\documentclass[aps,showpacs,twocolumn,superscriptaddress,amsmath,amssymb]{revtex4}

\usepackage{hyperref}
\usepackage{graphicx,dcolumn,bm}
\usepackage{amsmath}
\usepackage{amsthm}
\usepackage{amsfonts}

\begin{document}

\title{Generation of Total Angular Momentum Eigenstates in Remote Qubits}

\author{A. Maser}
\affiliation{Institut f\"ur Optik, Information und Photonik, Max-Planck Forschungsgruppe,\\Universit\"at Erlangen-N\"urnberg, 91058 Erlangen, Germany}

\author{U. Schilling}
\affiliation{Institut f\"ur Optik, Information und Photonik, Max-Planck Forschungsgruppe,\\Universit\"at Erlangen-N\"urnberg, 91058 Erlangen, Germany}

\author{T. Bastin}
\affiliation{Institut de Physique Nucl\'eaire, Atomique et de Spectroscopie, Universit\'e de Li\`ege, 4000 Li\`ege, Belgium}

\author{E. Solano}
\affiliation{Departamento de Qu\'imica F\'isica, Universidad del Pa\'is Vasco - Euskal Herriko Unibertsitatea, Apartado 644, 48080 Bilbao, Spain} 

\author{C. Thiel}
\email{cthiel@optik.uni-erlangen.de}
\homepage{http://www.ioip.mpg.de/jvz/}
\affiliation{Institut f\"ur Optik, Information und Photonik, Max-Planck Forschungsgruppe,\\Universit\"at Erlangen-N\"urnberg, 91058 Erlangen, Germany}

\author{J. von Zanthier}
\affiliation{Institut f\"ur Optik, Information und Photonik, Max-Planck Forschungsgruppe,\\Universit\"at Erlangen-N\"urnberg, 91058 Erlangen, Germany}

\date{\today}

\begin{abstract}
We propose a scheme enabling the universal coupling of angular momentum of $N$ remote noninteracting qubits using linear optical tools only. Our system consists of $N$ single-photon emitters in a $\Lambda$-configuration that are entangled among their long-lived ground-state qubits through suitably designed measurements of the emitted photons. In this manner, we present an experimentally feasible algorithm that is able to generate any of the $2^N$ symmetric and nonsymmetric total angular momentum eigenstates spanning the Hilbert space of the $N$-qubit compound.
\end{abstract}

\pacs{42.50.Dv,42.50.Tx,37.10.-x,03.67.-a}

\maketitle

\section{Introduction}

Since the celebrated article by Einstein, Podolsky, and Rosen in 1935~\cite{Einstein:1935:a}, it is commonly assumed that the phenomenon of entanglement between different systems occurs if the systems {\em had previously interacted with each other}. Indeed, for most experiments generating entangled quantum states interactions such as non-linear effects~\cite{Kwiat:1995:a}, atomic collisions~\cite{Osnaghi:2001:a}, Coulomb coupling~\cite{Leibfried:2005:a,Haeffner:2005:a}, or atom-photon interfaces~\cite{Wilk:2007:a}, are a prerequisite. Recent proposals considered that entanglement between systems that never interacted before can be created as a consequence of measuring photons propagating along multiple quantum paths, leaving the emitters in particular entangled states~\cite{Cabrillo:1999:a,Bose:1999:a,Skornia:2001:a,Duan:2001:a,Feng:2003:a,Duan:2003:a,Simon:2003:a,Thiel:2007:a,Bastin:2007:a}. Since then, several experiments generating {\em entanglement at a distance via projection} have been realized, first between disordered clouds of atoms~\cite{Julsgaard:2001:a,Chou:2005:a,Matsukevich:2006:a} and very recently even between single trapped atoms~\cite{Moehring:2007:a}.

On the other hand, the coupling of angular momentum is commonly utilized to account for the interaction between particles in order to retrieve the corresponding energy eigenstates and eigenvalues of the total system. This coupling of angular momentum has been fruitfully employed in as disparate fields as solid state, atomic or high-energy physics, to account for the interaction between electric or magnetic multipoles or spins of quarks, respectively~\cite{Wigner:1959}. Here again, it seems counter-intuitive that noninteracting particles, such as remotely placed spin-1/2 particles, will couple to form arbitrary total angular momentum eigenstates as if an interaction were present, including highly and weakly entangled quantum states.

In this article, we propose a method how to mimic the universal coupling of angular momentum of $N$ remote noninteracting spin-1/2 particles (qubits) in an experimentally operational manner. Hereby, an arbitrary number of distant particles can be entangled in their two-level ground states providing long-lived $N$-qubit states via the use of suitably designed projective measurements. In reference to the algorithm describing the coupling of angular momentum of individual spin-1/2 particles, our method couples successively remote qubit states to a multi-qubit compound system. Thereby, it offers access to the entire coupled basis of an $N$-qubit compound system of dimension $2^N$, i.e., to any of the $2^N$ symmetric and nonsymmetric total angular momentum eigenstates.

\section{Description of the physical system}

For $N$ spin-1/2 particles, the total angular momentum eigenstates, defined as simultaneous eigenstates of the square of the total spin operator $\hat{\bf S}^2$ and its $z$-component $\hat{S}_z$, are commonly denoted by $|S_N;\!m_N\rangle$, with the corresponding eigenvalues $S_N(S_N+1)\hbar^2$ and $m_N\hbar$~\cite{Dicke:1954:a,Mandel:1995:a}. However, since the denomination $|S_N;\!m_N\rangle$ generally characterizes more than one quantum state, we will extend the notation of an $N$-qubit state by its coupling history, i.e.~by adding the values of $S_1, S_2, ..., S_{N-1}$ to those of $S_N$ and $m_N$. A single qubit state has $S_1=\frac{1}{2}$, a two-qubit system can either have $S_2=0$ or $S_2=1$, a three-qubit system $S_3=\frac{1}{2}$ or $S_3=\frac{3}{2}$, and so on. Including the coupling history we thus get the following notation $|S_1,\!S_2,...,\!S_N;\!m_N\rangle$ which describes a particular angular momentum eigenstate unambiguously.

\begin{figure}[t!]
\centering
\includegraphics[width=0.45\textwidth, bb=100 580 485 755, clip=true]{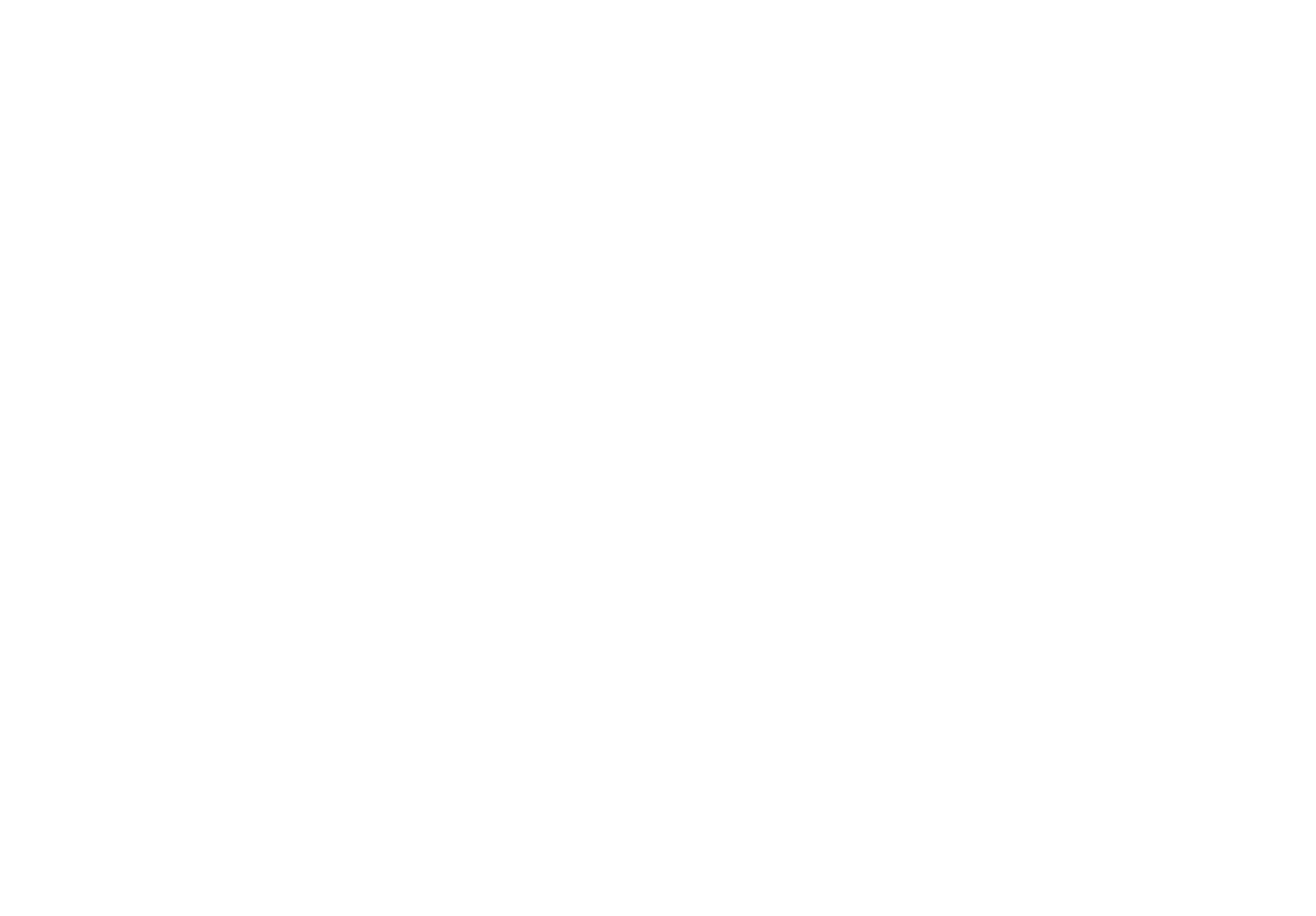}
\caption{\label{fig1} (Color online) Experimental setup for the angular momentum coupling of two atoms via projective measurements using optical fibers. In a successful measurement cycle, each atom emits a single photon and each detector registers exactly one photon. Note that the detectors cannot distinguish which of the atoms emitted a registered photon.}
\end{figure}

In the following, we consider a system consisting of $N$ indistinguishable single-photon emitters, e.g.~atoms, with a $\Lambda$-configuration, see Fig.~\ref{fig1}. We denote the two ground levels of the $\Lambda$-configured atoms as $|+\rangle$ and $|-\rangle$ or, using the notation introduced before, $|\frac{1}{2};\!+\frac{1}{2}\rangle\equiv|+\rangle$ and $|\frac{1}{2};\!-\frac{1}{2}\rangle\equiv|-\rangle$. Initially, all atoms are excited by a laser $\pi$ pulse towards the excited state $|e\rangle$ and subsequently decay by spontaneously emitting $N$ photons that are collected by single-mode optical fibers~\cite{Moehring:2007:a,Volz:2006:a} and transmitted to $N$ different detectors. Since each atom is connected via optical fibers to several detectors, a single photon can travel on several alternative, yet equally probable paths to be eventually recorded by one detector. After a successful measurement, where all $N$ photons have been recorded at the $N$ detectors so that each detector registers exactly one photon, it is thus impossible to determine along {\em which way} each of the $N$ photons propagated. This may cause quantum interferences of $N$th order which can be fruitfully employed to engineer particular quantum states of the emitters, e.g., to generate families of entangled states symmetric under permutation of their qubits~\cite{Thiel:2007:a,Bastin:2007:a}. Here, we will consider the generation of a more general class of quantum states, including symmetric {\em and} nonsymmetric states. By mimicking the process of spin-spin coupling, we will demonstrate how to generate any quantum state belonging to the coupled basis of an $N$-qubit compound system.

\section{Measurement based preparation of total angular momentum eigenstates}

Let us start by looking at the most basic process of our system. If one single excited atom with a $\Lambda$-configuration emits a photon, the atomic ground state and the photonic polarization states cannot be described independently. The excited state $|e\rangle$ can decay along two possible channels, $|e\rangle\rightarrow|+\rangle$ and $|e\rangle\rightarrow|-\rangle$, accompanied by the spontaneous emission of a $\sigma^-$ or a $\sigma^+$-polarized photon, respectively (consider e.g.~Zeeman sub-levels). A single decaying atom thus forms an entangled state between the polarization state of the emitted photon and the corresponding ground state of the de-excited atom~\cite{Volz:2006:a,Blinov:2004:a}. This correlation implies that the state of the atom is projected onto $|+\rangle$ ($|-\rangle$) if the emitted photon is registered by a detector with a $\sigma^-$ ($\sigma^+$) polarized filter in front.

\subsection{Preparation of $2$-qubit states}

In a next step, we consider the system shown in Fig.~\ref{fig1} where two atoms with a $\Lambda$-configuration are initially excited and subsequent measurements on the spontaneously emitted photons are performed at two different detectors. Again, if a polarization sensitive measurement is performed on the two emitted photons using two different polarization filters in front of the detectors, the state of the two atoms is projected due to the measurement. However, if the polarization of both photons is measured along orthogonal directions, the state of the atoms will be projected onto a superposition of both ground states, since it is impossible to determine which atom emitted the photon travelling to the first or the second detector by the information obtained in the measurement process. With each qubit having a total spin of $\frac{1}{2}$, a two-qubit system can have a total spin of either $1$ or $0$ and thus defines four angular momentum eigenstates given by:
\begin{tabular*}{\textwidth}{cccccc}\\
spin-1 triplet & $|S_1,\!S_2;\!m\rangle$ && spin-0 singlet & $|S_1,\!S_2;\!m\rangle$ & \vspace{3mm}\\
$|\!\!+\!+\!\rangle$ & $|\frac{1}{2},\!1;\!+1\rangle$ &&&& \vspace{1mm}\\ 
$\frac{1}{\sqrt{2}}(|\!\!+\!-\!\rangle\!+\!|\!\!-\!+\!\rangle)$ & $|\frac{1}{2},\!1;\!0\rangle$ && $\frac{1}{\sqrt{2}}(|\!\!+\!-\!\rangle\!-\!|\!\!-\!+\!\rangle)$ & $|\frac{1}{2},\!0;\!0\rangle$ \vspace{1mm}\\
$|\!\!-\!-\!\rangle$ & $|\frac{1}{2},\!1;\!-1\rangle$ &&&& \vspace{3mm}
\end{tabular*}
The spin-1 triplet can be easily generated with the setup shown in Fig.~\ref{fig1} by choosing the polarization filters accordingly: For example, if both filters are oriented in such a way that only $\sigma^-$ ($\sigma^+$) polarized photons are transmitted, the emitters are projected onto the state $|\!\!+\!+\rangle$ ($|\!\!-\!-\rangle$); if the filters are orthogonal, i.e.~one is transmitting $\sigma^-$ and one $\sigma^+$ polarized photons, the system is projected onto the state $|\frac{1}{2},\!1;\!0\rangle$, since any information along {\em which way} the photons propagated is erased by the system. Finally, in order to generate the singlet state $|\frac{1}{2},\!0;\!0\rangle$, we may introduce an optical phase shift of $\pi$ in one of the optical paths shown in Fig.~\ref{fig1}, e.g., by extending or shortening the length of the optical path by $\frac{\lambda}{2}$. The generation of the four two-particle total angular momentum eigenstates with the system shown in Fig.~\ref{fig1} thus requires only the variation of two polarizer orientations and, in case of the singlet state, to introduce an optical phase shift of $\pi$. 

\subsection{Preparation of $3$-qubit states}

With the two-qubit angular momentum eigenstates at hand, we can next couple an additional qubit in order to access the eight possible three-qubit total angular momentum eigenstates. In the following, we will exemplify our method for the three-qubit state $|\!\frac{1}{2},\!1,\!\frac{1}{2};\!+\frac{1}{2}\rangle$ given by
\begin{eqnarray}\label{3Dicke}\textstyle
|\frac{1}{2},\!1,\!\frac{1}{2};\!+\frac{1}{2}\rangle&=&\frac{1}{\sqrt{6}}\,(2|\!+\!+-\rangle-|\!+\!-+\rangle-|\!-\!++\rangle)\\
&=&\frac{\sqrt{2}}{\sqrt{3}}\,|\frac{1}{2},\!1;+\!1\rangle\otimes|-\rangle-\frac{1}{\sqrt{3}}\,|\frac{1}{2},\!1;\!0\rangle\otimes|+\rangle\nonumber,
\end{eqnarray}
where the last line in Eq.~(\ref{3Dicke}) exhibits the coupling history: In order to generate the three-qubit state $|\frac{1}{2},\!1,\!\frac{1}{2};\!+\frac{1}{2}\rangle$, the two-qubit spin-1 states $|\frac{1}{2},\!1;\!+1\rangle$ and $|\frac{1}{2},\!1;\!0\rangle$ are coupled with $|-\rangle$ and $|+\rangle$, respectively. Thereby, the prefactors $\frac{\sqrt{2}}{\sqrt{3}}$ and $-\frac{\sqrt{1}}{\sqrt{3}}$ represent the corresponding Clebsch-Gordan coefficients as a result of changing the basis~\cite{Clebsche:2002:a}. In the following, we will make use of our knowledge of how to generate the states $|\frac{1}{2},\!1;\!+1\rangle$ and $|\frac{1}{2},\!1;\!0\rangle$ in order to generate the desired state $|\frac{1}{2},\!1,\!\frac{1}{2};\!+\frac{1}{2}\rangle$. Therefore, we have to add a third qubit and combine the two systems generating the two individual states accordingly in one setup.

\begin{figure}[t!]
\begin{center}
\includegraphics[width=0.23\textwidth, bb=65 500 460 810, clip=true]{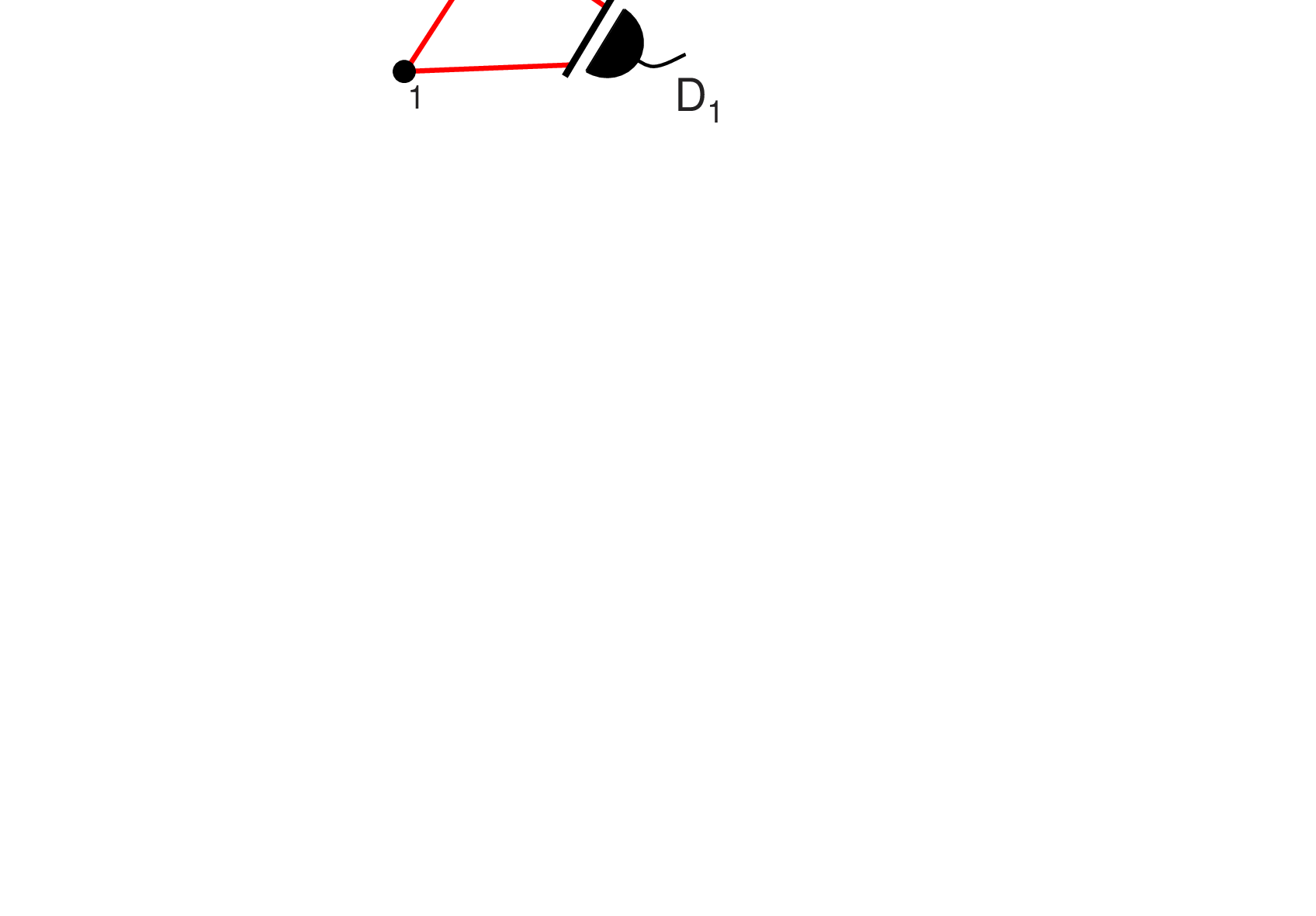}
\includegraphics[width=0.23\textwidth, bb=65 500 460 810, clip=true]{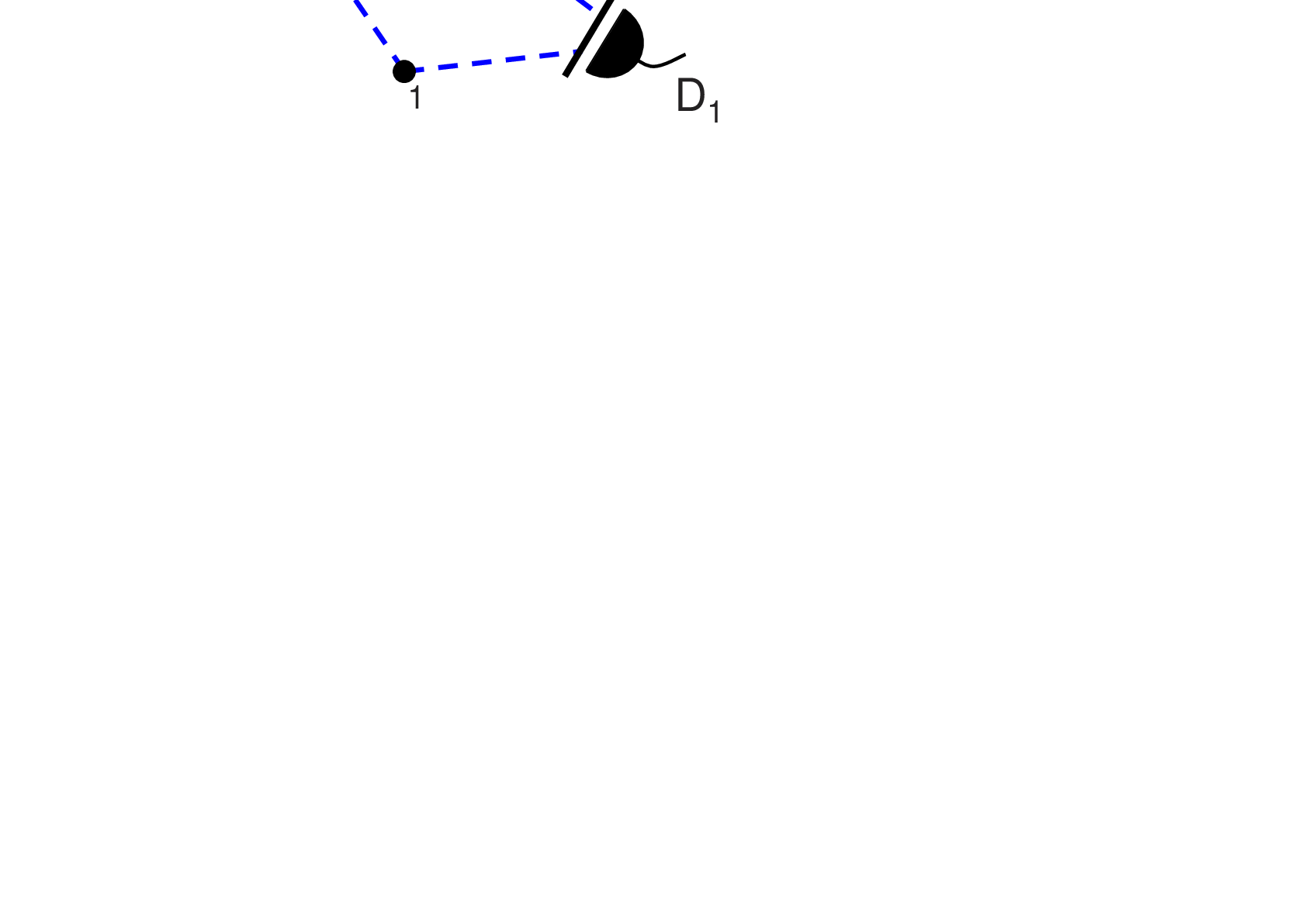}
\caption{\label{fig2}(Color online) Left: Extension of the setup shown in Fig.~\ref{fig1} capable of generating the state $|\frac{1}{2},\!1;\!0\rangle\otimes|+\rangle$. Right: configuration for the generation of the state $|\frac{1}{2},\!1;\!1\rangle\otimes|-\rangle$.}
\end{center}
\end{figure}

\begin{figure}[t!]
\begin{center}
\includegraphics[width=0.3\textwidth, bb=165 485 535 780, clip=true]{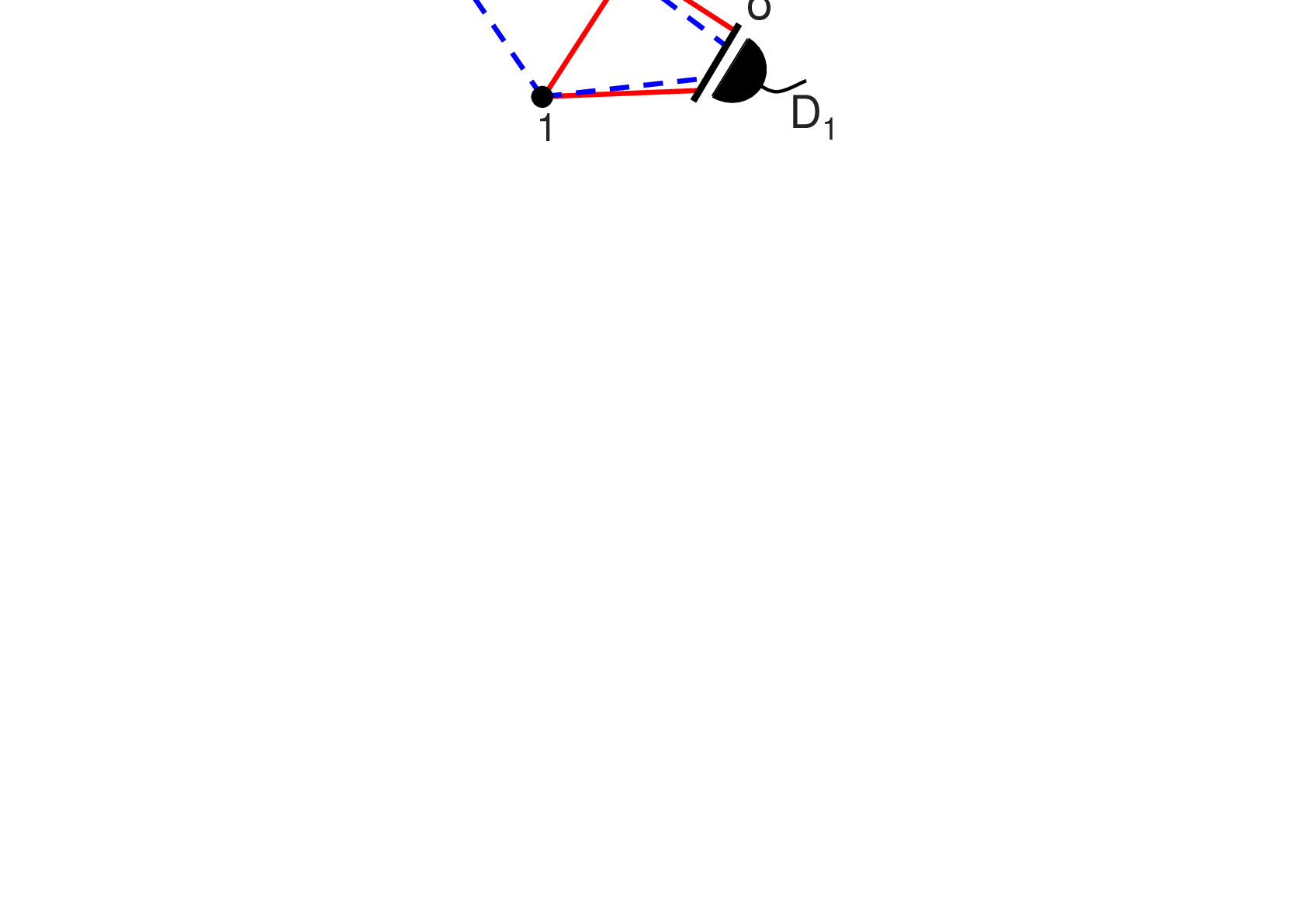}
\caption{(Color online) Setup for the generation of the state $2|{+\!+\!-}\rangle-|{+\!-\!+}\rangle-|{-\!+\!+}\rangle$. The blue dashed lines indicate the quantum path which leads to $2|{+\!+\!-}\rangle$, whereas the red solid labeled path leads to $-|{+\!-\!+}\rangle-|{-\!+\!+}\rangle$. Please note that the different red solid and blue dashed lines leading from atom 1 (2) to detector~$D_1$ are drawn to indicate the different quantum paths only. Physically, there is only one fiber from atom 1 (2) to detector~$D_1$.\label{fig3}}
\end{center}
\end{figure}

The two setups individually capable of generating the three-qubit states $|\frac{1}{2},\!1;\!+1\rangle\otimes|-\rangle$ and $|\frac{1}{2},\!1;\!0\rangle\otimes|+\rangle$ are shown in Fig.~\ref{fig2}. The additional qubit is not yet coupled to the two-qubit system, i.e.~it is simply projected either onto the state $|+\rangle$ (Fig.~\ref{fig2}, left) or $|-\rangle$ (Fig.~\ref{fig2}, right), where the two-qubit systems are projected in the same way as explained in Fig.~\ref{fig1}. In order to generate the three-qubit state $|\frac{1}{2},\!1,\!\frac{1}{2};\!\frac{1}{2}\rangle$, we now have to superpose these two possibilities. The combined system is shown in Fig.~\ref{fig3}. We will explain the underlying physics by considering the possible scenarios when detecting the photon emitted by the additional third atom.

In a successful measurement cycle, the three emitted photons are detected at three different detectors. Thus, there are only two possible situations due to a measurement of a photon emitted by the third atom:
\begin{itemize}
\item[I.] (red solid lines) The emitted photon is registered at detector~$D_3$ which has a $\sigma^-$ polarizing filter in front. In this case, emitter 3 is projected onto the state $|+\rangle$ and emitter 1 and 2 are left in the setup generating the state $|\frac{1}{2},\!1;\!0\rangle\equiv\frac{1}{\sqrt{2}}(|\!+\!-\rangle+|\!-\!+\rangle)$; as discussed in Fig.~\ref{fig2} (left).
\item[II.] (blue dashed lines) The emitted photon is registered at detector~$D_2$ which has a $\sigma^+$ polarizing filter in front. In this case, emitter 3 is projected onto the state $|-\rangle$ and emitter 1 and 2 are left in the setup generating the state $|\frac{1}{2},\!1;\!1\rangle\equiv|\!+\!+\rangle$; as discussed in Fig.~\ref{fig2} (right).
\end{itemize}
In other words, the third emitter acts as a switch between the two possible quantum paths: with equal probabilities, the system is either projected onto the state $2|++-\rangle$ or onto the state $|+-+\rangle+|-++\rangle$. Note that the relative factor of two results from using an equal number of path ways (optical fibers) in both cases. In addition, we can modify the path where a photon emitted by the third atom is registered at detector~$D_3$ by implementing a relative optical phase shift of $\pi$ (c.f.~Fig.~\ref{fig3}) to obtain a minus sign for scenario II. relative to scenario I. In this case, the final state projected by the setup shown in Fig.~\ref{fig3} corresponds to the three-qubit state $|\frac{1}{2},\!1,\!\frac{1}{2};\!\frac{1}{2}\rangle$ of Eq.~(\ref{3Dicke}).

Reconsidering the state $|\frac{1}{2},\!1,\!\frac{1}{2};\!\frac{1}{2}\rangle$ in terms of our extended notation, we coupled two spin-1/2 particles to form a spin-1 compound state that was coupled again with a spin-1/2 particle to form a three-particle spin-1/2 compound state. Similarly, we could have coupled the spin-1 compound state with an additional qubit in such a way that we obtain the symmetric state $|\frac{1}{2},\!1,\!\frac{3}{2};\!\frac{1}{2}\rangle$, also known as W-state~\cite{Duer:2000:a}. For this case, we have to change the setup shown in Fig.~\ref{fig3} slightly: we remove the optical phase shift of $\pi$ and connect the third emitter also with detector~$D_1$. In this case, the totally symmetric setup generates a W-state (c.f.~\cite{Thiel:2007:a}).

\begin{figure}[b!]
\centering
\includegraphics[width=0.41\textwidth, bb=240 325 620 760, clip=true]{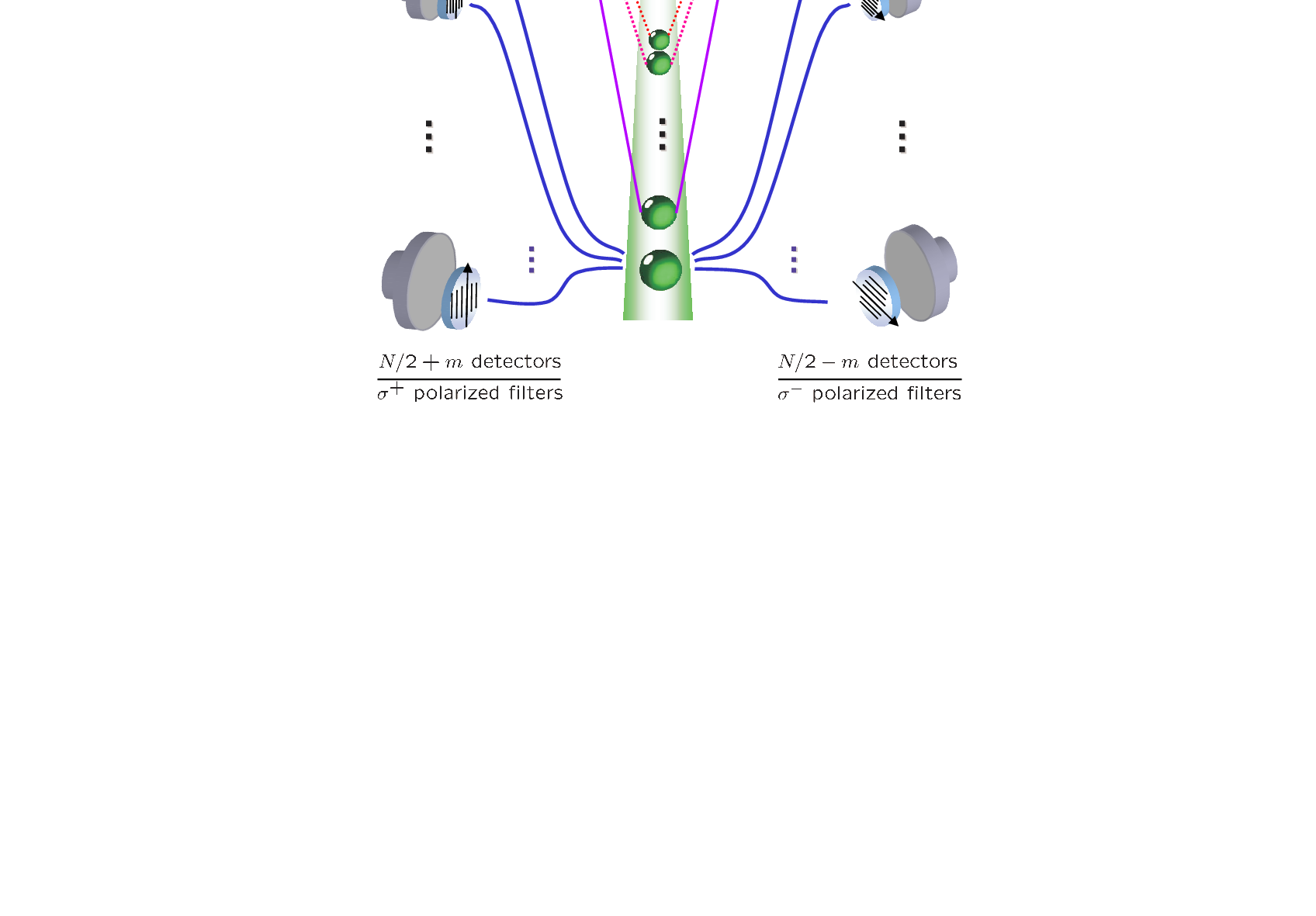}
\caption{\label{fig4} (Color online) Experimental setup for the spin-spin coupling of $N$ remote atoms via projective measurements.}
\end{figure}

\subsection{Preparation of $N$-qubit states}

Finally, let us outline how to engineer the coupling of angular momentum of $N$ remote qubits to form an arbitrary $N$-qubit total angular momentum eigenstate. In order to generate the $N$-qubit state $|{S_1,S_2,S_3,...S_N;m_s}\rangle$ we have to 
\begin{itemize}
\item[1.] set up $\frac{N}{2}+m_s$ ($\frac{N}{2}-m_s$) detectors with $\sigma^-$ ($\sigma^+$) polarized filters in front. Hereby, we connect the first emitter with optical fibers to all $N$ detectors.
\item[2.] check for each particle $i$ beginning with $i=2$ whether $S_i>S_{i-1}$ or $S_i<S_{i-1}$. If
\begin{itemize}
\item[a.] $S_i>S_{i-1}$; we have to connect the particle with optical fibers to all detectors except those which are mentioned in case~b.~below.
\item[b.] $S_i<S_{i-1}$; we have to connect the particle with optical fibers to one detector with a $\sigma^-$ polarizer and to one with a $\sigma^+$ polarizer. The optical fiber leading to the $\sigma^-$ polarizer should induce a relative optical phase shift of $\pi$ and those two detectors should not be linked to any other subsequent particle.
\end{itemize}
\end{itemize}
If one wants to create a particular total angular momentum eigenstate $|{S_1,S_2,S_3,...S_N;m_s}\rangle$, the setup is determined by the total spins $S_1,S_2,S_3,...S_N$ obtained by successively coupling $N$ spin-1/2 particles. Hereby, the spin number $m_s$ determines the fraction of $\sigma^-$ and $\sigma^+$ polarized filters used in the setup (s.~Fig.~\ref{fig4}).

As examples, let us apply this algorithm for the two three-qubit total angular momentum eigenstates $|\frac{1}{2},\!1,\!\frac{1}{2};\!\frac{1}{2}\rangle$ and $|\frac{1}{2},\!1,\!\frac{3}{2};\!\frac{1}{2}\rangle$ discussed above. Since $m_s=\frac{1}{2}$ for both states, we use two detectors with $\sigma^-$ polarized filters and one with a $\sigma^+$ polarized filter. Further, in both cases we have $S_2>S_1$ which implies that the first and the second emitter are connected to all three detectors. For the state $|\frac{1}{2},\!1,\!\frac{1}{2};\!\frac{1}{2}\rangle$, we find $S_3<S_2$. Therefore, we connect the third emitter only to two detectors with $\sigma^-$ and $\sigma^+$ polarized filters in front, respectively, e.g.~detector~$D_2$ and~$D_3$, and we introduce an optical phase shift of $\pi$ for the path leading from the third emitter to detector~$D_3$. Summarizing we obtain the setup shown in Fig.~\ref{fig3} as postulated. For the state $|\frac{1}{2},\!1,\!\frac{3}{2};\!\frac{1}{2}\rangle$, we find $S_3>S_2$. Here, we connect the third emitter to all three detectors. In this case, as mentioned above, the setup will generate the symmetric W-state~\cite{Thiel:2007:a}.

The method proposed here, relies on the probabilistic scattering of photons. Thereby, the efficiency of generating a particular $N$-qubit total angular momentum eigenstate decreases with increasing number of qubits $N$. If the probability to find a single photon in an angular detection window $\Delta\Omega$ is given by $P(\Delta\Omega)$, including fiber coupling and detection efficiencies, the corresponding $N$-fold counting rate is found to be $P^N(\Delta\Omega)$. This might limit the scalability of our scheme (see the discussion in~\cite{Thiel:2007:a}) as is indeed the case with other experiments observing entangled atoms~\cite{Blinov:2004:a,Volz:2006:a,Moehring:2007:a}.

\section{Conclusions}

In conclusion, we considered a system of $N$ remote noninteracting single-photon emitters with a $\Lambda$-configuration. By mimicking the coupling of angular momentum, we showed that it is possible to engineer any of the $2^N$ total angular momentum eigenstates in the long-lived ground-state qubits. Using linear optical tools only, our method employs the detection of all $N$ photons scattered from the $N$ emitters at $N$ polarization sensitive detectors. Thereby, it offers access to any of the $2^N$ states of the coupled basis of an $N$-qubit compound system. Using projective measurements we thereby form highly and weakly entangled quantum states even though no interaction between the qubits is present.

\section{Acknowledment}

U.S. thanks financial support from the Elite Network of Bavaria. E.S. thanks financial support from Ikerbasque Foundation, EU EuroSQIP project, and UPV-EHU Grant GIU07/40. C.T.~and J.v.Z.~gratefully acknowledge financial support by the Staedtler foundation.

\end{document}